\documentclass[12pt]{article}
\usepackage{amsmath}
\usepackage{graphicx}
\usepackage{amssymb}
\usepackage{amsfonts}
\usepackage{latexsym}

\def\beq{\begin{eqnarray}}
\def\eeq{\end{eqnarray}}

\def\det{\,\mbox{det}\,}

\def\pa{\partial}

\def\si{\sigma}

\newcommand {\dl}   {\delta}

\newcommand {\lm}   {\lambda}      
          
\newcommand {\s }   {\sigma}       
         
\newcommand {\vf }  {\varphi}

\newcommand {\pl}   {\partial}     

\newcommand   {\diag}{{\sf\,diag\,}}

\newcommand {\vol}  {\sqrt{|g|}}

\newcommand {\MR}  {{\mathbb R}}

\begin{document}

\begin{center}
{\large\bf Schr{\"{o}}dinger equation in the space with 
cylindrical geometric defect and possible application 
to multi-wall nanotubes}
\vskip 4mm

{\bf Guilherne de Berredo-Peixoto$^{(a)}$, 
Mikhail O. Katanaev$^{(b)}$, 
\\
Elena Konstantinova$^{(a,c)}$ and 
Ilya L. Shapiro$^{(a)}$} 
\vskip 4mm 

{\small \sl 
(a) \ \ Departamento de F\'{\i}sica, 
           Universidade Federal de Juiz de Fora, \\
           \sl Juiz de Fora, CEP 36036--330, MG, Brazil}
\vskip 4mm 

{\small  (b) \ \ 
Steklov Mathematical Institute, 
           \sl Gubkin St. 8, Moscow, 119991, Russia}
\vskip 4mm 

{\small \sl (c) \ \ 
Instituto Federal de Educa\c{c}\~ao, Ci\'{e}ncia e Tecnologia do 
Sudeste de
Minas Gerais (IFSEMG), Juiz de Fora, CEP 36080-001, MG, Brazil}
\end{center}

\vskip 12mm 

\centerline{\bf Abstract}
\vskip 2mm 

The recently invented cylindrical geometric space defect 
is applied to the electron behaviour in the system which 
can be regarded as a simplified model of a double-wall 
nanotube. By solving the Schr$\ddot{\rm o}$dinger equation 
in the region of space with cylindrical geometric defect 
we explore the influence of such geometric defect on the 
energy gap and charge distribution. The effect is 
qualitatively similar to the one obtained earlier by 
means of traditional simulation methods. In general, the 
geometric approach can not compete with the known methods 
of theoretical study of the nanostructures, such as 
molecular dynamics. However it may be useful for better 
qualitative understanding of the electronic properties of 
the nanosystems.
\vskip 6mm

\section{Introduction}

The theory of defects is an important part of the mathematical 
physics. In particular, topological defects attract much 
attention in modern condensed matter physics (see, e.g., 
\cite{ChaiLub,defects} for an introduction and recent review). 
Another elegant approach in this area is the geometric theory 
of defects \cite{KatVol92,KatVol99} (see also \cite{Katana05}
for the introduction), which is formulated in terms 
of the notions originally developed in the theories 
of gravity. The main kinds of defects which were described 
in the framework of Riemann–-Cartan geometry are dislocations 
and disclinations. This means that the curvature and torsion 
tensors are interpreted as surface densities of Frank and 
Burgers vectors and thus linked to the nonlinear, generally 
inelastic deformations of a solid. Recently, the qualitatively 
new kind of geometric defect corresponding to the cylindrical 
geometry has been described in \cite{deBKat09}. 
In the framework of geometric approach, the static massive 
thin cylindrical shells are considered as sources 
in Einstein equations. It turns out that these defects 
correspond to $\delta$- and $\delta^\prime$--function type 
energy-momentum tensors. It would be interesting to find some 
useful applications of this new kind of defect. One may 
think about cosmological applications and/or about the 
condensed matter physics applications, which can be not 
infinitely far from each other \cite{Cosm-Cond}.

One of the natural question to address is in which way the 
new kind of defects can be applied to the condensed matter 
physics? In this respect we can mention, for example, 
the recent works \cite{Das,Cheung} devoted to the 
cylindrical nanoparticles in a liquid crystal solvent. 
It would be interesting to see whether the cylindric 
defects can be useful in description of these systems.  
However, at the first place, the cylindrical geometry is 
associated with the nanotubes (see, e.g., 
\cite{nanotube,exper,Applic1,Applic2} for the general review). 
As we shall see in what follows, the most appropriate object 
of our interest would be not the usual single-wall nanotubes 
but the double-wall nanotubes (DWNTs) or, more general, 
multi-wall nanotubes (MWNTs). 

The MWNTs attract great attention \cite{MWCNT,gap2,chir}, 
in particular due to higher temperature stability and 
stiffness compared to the similar single-wall nanotubes.
The inner and the outer 
layers of the DWNTs can be metallic (M) or semiconductor (S). 
Correspondingly, there are DWNTs of different kinds, namely: 
semiconductor-semiconductor (S/S case), metal-semiconductor 
(M/S case),  semiconductor-metal (S/M case) and  metal-metal 
(M/M case). 
It has been emphasized recently in \cite{today} that the 
M/M type of the DWNTs is the most difficult to meet, but, 
at the same time, this particular configuration is especially 
interesting  to explore. Let us note that there are some 
theoretical and experimental data available on the energy 
gap of a MWNTs compared to the purely metallic tubes 
\cite{gap} (see also \cite{gap2}). 

In the present work we shall consider how the cylindrical 
geometric defects can be related to the theoretical 
investigation of the M/M type DWNTs. Our approach will 
be to consider the solution of the one-particle 
Schr$\ddot{\rm o}$dinger equation 
in the space with the cylindrical geometric defect. Let 
us note that the idea to explore the one-particle 
Schr$\ddot{\rm o}$dinger equation as a way to 
describe nanotubes is not completely new. One 
can mention, e.g., the work of Ref. \cite{GraWill}, 
where a nonrelativistic spinless particle moving 
in a cylindrical surface (also in a thick 
cylindrical hollow) is investigated. On the other 
side, there are also publications on quantized particles 
moving in a curved background, with or without defects. 
Many of them address the problem of a test particle 
in a spacetime with cosmic strings and are motivated 
by the corresponding cosmological models. Also, in 
Ref. \cite{LongShore}, for example, the 
Schr$\ddot{\rm o}$dinger picture description of 
vacuum states is studied and applied to simple 
cosmological models without cosmic strings. 

An important difference between the works mentioned 
above and our approach is that we do not consider the 
cylindrically symmetric potential. Instead, we consider the 
free Schr$\ddot{\rm o}$dinger equation (without potential) 
in the space with the cylindrical geometric defect. In order 
to see in which situation this method may be more natural, 
let us present the following observations. The conventional 
nanotubes are compounds where the electron can propagate 
only on the tube surface and not in the three-dimensional 
bulk. In the case of DWNTs there are two distinct conducting 
cylindrical shells. If we consider the electron in such 
system, the tube or space division shell between the two 
tubes is unlikely to create an electromagnetic potential 
difference between the two conducting layers while being 
essentiall for the electronic properties of the compound. 
Therefore, the presence of the division between the two 
shells fits the geometric defect case described in 
\cite{deBKat09}, so it looks natural to perform some 
investigation of the corresponding quantum systems. 

It is necessary to mention some recent publications 
motivated by condensed matter 
applications of geometric approach. In Ref. \cite{mosta}, 
the effect of curvature is analyzed in particle scattering. 
Concerning the problems in the presence of defects, we can 
also mention, e.g., Ref. \cite{azevedo}, where a particle 
which moves in a magnetic field in a space with disclination
and screw dislocation has been studied. Also, the paper 
\cite{furtado} explores the bounded states of an electric 
dipole in the presence of a conical defect. Despite our 
work can be seen as continuation of this line, we consider 
a rather different approach to the geometrical defects
\cite{deBKat09}.  

Indeed, there 
are well-known methods for calculating electronic 
and mechanic properties for different compounds, $3D$, 
$2D$ or $1D$ nanosystems, in particular. For example, 
using the density functional (DFT) - based methods we 
can explore different systems (including periodical ones), 
and one obtains the desirable results with high precision
\cite{DFT}. At the same time, it looks interesting to 
have an alternative, more simple (albeit potentially 
reliable) approach which could permit us to obtain 
important qualitative information, and maybe also 
help in qualitative understanding, e.g., of the electronic
properties of the mentioned systems. 
In what follows we shall develop relatively 
simple and to great extent analytic method based on 
the geometric theory of defects. As we shall see in 
brief, this approach looks justified, for it enables 
one to arrive at the qualitative understanding of the 
origin of modifications of energy gaps and density 
distribution due to the presence of the defect shell 
between the conducting layers.

The paper 
is organized as follows. In the next two sections we present 
a brief but to some extent pedagogical description of 
the tube dislocation in the linear elasticity theory
and in the geometric theory of defects. Despite the 
content of these sections is essentially the same as 
the one of \cite{deBKat09}, we include it here com the 
sake of completeness. In section 4 we consider the 
Schr$\ddot{\rm o}$dinger 
equation and describe the results of its numerical solution. 
In particular, we compare the energy levels and charge 
density distributions of the MWNTs with the ones of the 
similar cylindrical tube without defect (it can be obtained 
analytically) and with the established earlier features of 
the energy spectrum of the system \cite{gap2, gap}.  
Finally, in section 5 we draw our conclusions.


\section{Tube dislocation in the linear elasticity theory}

Let us start by describing the tube dislocations in the 
simplest case of linear elasticity theory. Consider a
homogeneous and isotropic 
elastic media as a three-dimensional Euclidean space 
$\MR^3$ with Cartesian coordinates $\,x^i,\,y^i$, 
where $\,i=1,2,3$. The Euclidean metric is denoted by
$\dl_{ij}=\diag(+++)$. The basic variable in the 
elasticity theory is the displacement vector of a 
point in the elastic media, $u^i(x)$, $\,x\in\MR^3$. 
In the absence of external forces, Newton's and Hooke's 
laws reduce to three second order partial differential
equations which describe the equilibrium state of elastic 
media (see, e.g.,~\cite{LanLif70}),
\begin{equation}                                                  
\label{eqsepu}
  (1-2\s)\triangle u_i+\pl_i\pl_j u^j=0\,.
\end{equation}
Here $\triangle$ is the Laplace operator and 
the dimensionless Poisson ratio $\si$ 
\ ($-1\le\s\le1/2$) \ is defined as 
\begin{equation*}
\s=\frac\lm{2(\lm+\mu)}\,.
\end{equation*}
$\lm$ and $\mu$ are called the Lame coefficients, they
characterize the elastic properties of media.

Raising and lowering of Latin indices $i,j,\dotsc$ 
can be done by using the Euclidean metric, $\dl_{ij}$, 
and its inverse, $\dl^{ij}$.
Eq. (\ref{eqsepu}) together with the corresponding 
boundary conditions enables one to establish the 
solution for the field  $u^i(x)$ in a unique way.

Let us pose the problem for the tube dislocation shown in 
Figure \ref{fig.1},{\it a}.

\begin{quotation}
\begin{figure}
\includegraphics{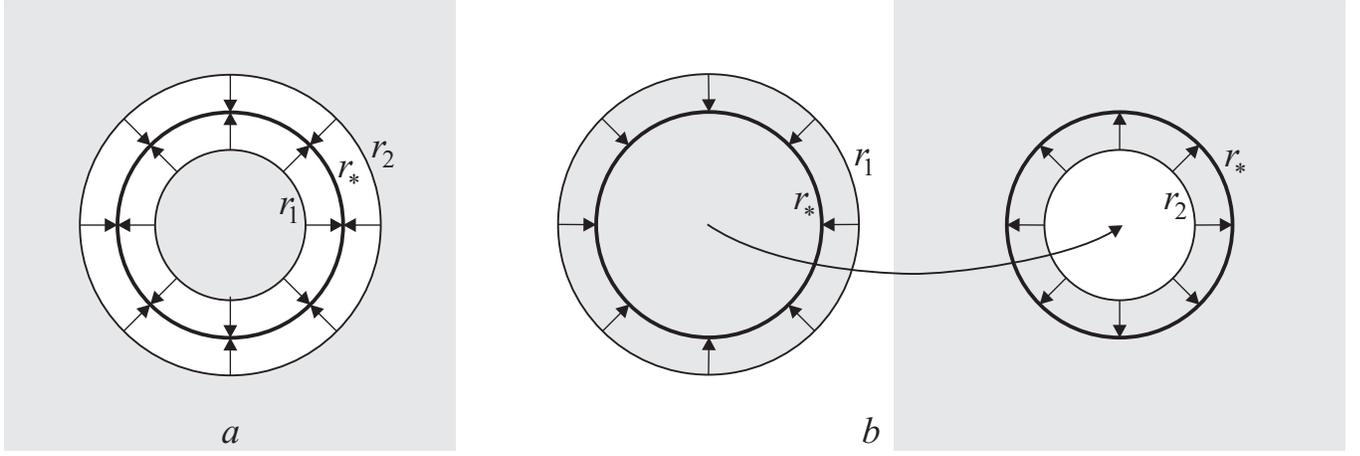}
\centering \caption{\label{fig.1} Negative ({\it a}) 
and positive ({\it b}) tube dislocations.}
\end{figure}
\end{quotation}
This tube dislocation can be produced as follows. We cut 
out the thick cylinder of media located between two parallel 
cylinders of radii $r_1$ and $r_2$ ($r_1<r_2$) with the axis 
$z=x^3$ as the symmetry axis of both cylinders, move 
symmetrically both cutting surfaces one to the other and 
finally glue them. Due to circular and translational
symmetries of the problem, in the equilibrium state the 
gluing surface is also the cylinder, of the radius $r_*$ 
which will to be found below.

Within the procedure described above and shown in 
Fig.\ref{fig.1},{\it a} we observe the negative tube 
dislocation because part of the media was removed. This 
corresponds to the case of $r_1<r_2$. 
However, the procedure can be applied in the opposite 
way by addition of extra media to $\MR^3$ as shown in 
Fig.\ref{fig.1},{\it b}. In
this case, we meet a positive tube dislocation and 
the inequality has an opposite sign, $r_1>r_2$.
It is important to note that the approach adopted here
above concerns our treatment of the division of the 
shells of the DWNT as a geometric defect. Of course, 
it can not be seen as a practical prescription for 
building nanotubes of other objects. 

Let us start by calculating the radius of the equilibrium 
configuration $r_*$. This problem is naturally formulated 
and solved in cylindrical coordinates $\,r,\,\vf,\,z$. Let 
us denote the displacement field components in these
coordinates by $\,u^r,\,u^\vf,\,u^z$. In our case, 
$\,u^\vf=u^z=0\,$ due to the symmetry of the problem, 
so that the radial displacement field $u^r(r)$ can be 
simply denoted as $\,u^r(r)=u(r)$.

The boundary conditions for the equilibrium tube 
dislocation are
\begin{equation}                                                  
\label{ebotud}
  u\big|_{r=0}=0\,,\qquad 
  u\big|_{r=\infty}=0\,,\qquad 
 \left. \frac {du_{\text{in}}}{dr}\right|_{r=r_*}
 =\left.\frac{du_{\text{ex}}}{dr}\right|_{r=r_*}\,.
\end{equation}
The first two conditions are purely geometrical, and the 
third one means the equality of normal elastic forces 
inside and outside the gluing surface in the equilibrium 
state. The subscripts ``in'' and ``ex'' denote the 
displacement vector field inside and outside the gluing 
surface, respectively.

Let us note that our definition of the displacement vector 
field follows \cite{Katana05}, but differs slightly from 
the one used in many other references. In our notations, 
the point with coordinates $y^i$, after elastic 
deformation, moves to the point with coordinates $x^i$:
\begin{equation}                                                  
\label{eeldef}
y^i\rightarrow x^i(y)=y^i+u^i(x)\,.
\end{equation}
The displacement vector field is the difference between 
new and old coordinates, $u^i(x)=x^i-y^i$. Indeed, we are
considering the components of the displacement vector field, 
$u^i(x)$, as functions of the final state coordinates of 
media points, $x^i$, while in other references they are 
functions of the initial coordinates, $y^i$. The two 
approaches are equivalent in the absence of dislocations 
because both sets of coordinates $x^i$ and $y^i$ cover 
the entire Euclidean space $\MR^3$. On the contrary, if 
dislocation is present, the final state coordinates $x^i$ 
cover the whole $\MR^3$ while the initial state coordinates
cover only part of the Euclidean space lying outside the 
thick cylinder which was removed. For this reason the final 
state coordinates represent the most useful choice here. 

The elasticity equations (\ref{eqsepu}) can be easily 
solved for the case of tube dislocation under consideration. 
The Laplacian and the divergence in the cylindrical
coordinates have the form
\begin{align}
\label{cyl}
\triangle u_r & = \frac1r\pl_r(r\pl_r u_r)
+\frac1{r^2}\pl^2_\vf u_r+\pl^2_z u_r
  -\frac1{r^2}u_r-\frac2{r^2}\pl_\vf u_\vf\,,
\\
\pl_i u^i&=\frac1r\pl_r(ru^r)+\frac1r\pl_\vf u^\vf
+\pl_z u^z\,,
\end{align}
where the indices are lowered using the Euclidean metric 
in cylindrical coordinates, 
$\,u_r=u^r$, $u_\vf=u^\vf r^2\,$ and $\,u_z=u^z$. 
Let us note that the last two terms in (\ref{cyl}) are 
due to the geometric covariant nature of the Laplace
operator, which is constucted on the basis of covariant
derivatives. 

One can remember that only the radial component differs 
from zero. The angular $\vf$ and $z$ components of 
equations (\ref{eqsepu}) are identically satisfied, and 
the radial component reduces to the ordinary differential 
equation,
\begin{equation}                                                  
\label{eradec}
  \pl_r\Big[\frac1r\pl_r(r u)\Big]\,=\,0\,,
\end{equation}
which has a general solution
\begin{equation*}
  u = ar-\frac{b}{r}\,,
\label{solu}
\end{equation*}
depending on the two arbitrary constants of integration 
$a$ and $b$. Due to the first two boundary conditions 
(\ref{ebotud}), the solutions inside and outside the 
gluing surface are
\beq
\label{elsopt}
\begin{aligned}
  u_{\text{in}} &=ar\,, \quad &a&>0,
\\
  u_{\text{ex}}&=-\frac br\,, \quad  &b&>0.
\end{aligned}
\eeq
The signs of the integration constants correspond to the 
negative tube dislocation shown in Fig.\ref{fig.1},{\it a}. 
For positive tube dislocation, Fig.\ref{fig.1},{\it b}, 
both integration constants have opposite signs, $a<0$
and $b<0$. 

Using the solution (\ref{solu}) and the third boundary 
condition (\ref{ebotud}), one can determine the radius 
of the gluing surface,
\begin{equation}                                                  
\label{eglsur}
  r_*^2=\frac ba.
\end{equation}
After simple algebra the integration constants can be 
expressed in terms of the initial radius of external and 
internal cylinders
\begin{equation}                                                  
\label{eintco}
a = \frac{r_2-r_1}{r_2+r_1}=\frac l{2r_*}\,,
\qquad
b=\frac{r_2^2-r_1^2}4=\frac{lr_*}2\,,
\end{equation}
where
\begin{equation*}
  l=r_2-r_1\,, \;\;\; {\rm and} \;\;\; r_*=\frac{r_2+r_1}2
\end{equation*}
are the thickness of the removed cylinder and the radius of 
the gluing surface, respectively. The first expression in 
(\ref{eintco}) restricts the range of the integration 
constant, $0<|a|<1$. 

It is interesting to note that the above formulas are 
applicable for both negative and positive tube dislocations.
One has to take $l>0$ and $l<0$ in these cases, respectively. 
Indeed, the negative thickness in the last case means that,
before the deformation, the external radius must be smaller 
than the internal one and that the proper deformation 
consists in inserting some matter between the cylindrical 
surfaces. In both cases we see that the gluing surface 
$r=r_*$ lies exactly in the middle between the radii $r_1$ 
and $r_2$. 

Finally, within the linear elasticity theory, 
Eq.(\ref{elsopt}) with the integration constants 
(\ref{eintco}) yields a complete solution for the tube 
dislocation. This solution is valid for small relative 
displacements, when \ $l/r_1\ll 1$ and $l/r_2\ll 1$.
It is remarkable that the solution obtained in the 
framework of linear elasticity theory does not depend 
on the Poisson ratio of the media. In this sense, the 
tube dislocation is a purely geometric defect which 
does not feel the elastic properties.

In order to use the geometric approach we have to present 
the results obtained above in other terms. For this end we 
compute the geometric quantities of  the manifold 
corresponding to the tube dislocation. From the geometric 
point of view, the elastic deformation (\ref{eeldef}) is 
a diffeomorphism between the given domains in the Euclidean 
space. The original elastic media $\MR^3$, before the 
dislocation is made, is described by Cartesian coordinates 
$y^i$ with the Euclidean metric $\dl_{ij}$. An inverse
diffeomorphism transformation $x\rightarrow y$ induces 
a nontrivial metric on $\MR^3$, corresponding to the tube 
dislocation. In Cartesian coordinates this metric has 
the form
\begin{equation*}
g_{ij}(x)\,=\,
\frac{\pl y^k}{\pl x^i}\,\frac{\pl y^l}{\pl x^j}\,\dl_{kl}.
\end{equation*}
We use curvilinear cylindrical coordinates for the tube 
dislocation and therefore it is useful to modify our 
notations. The indices in curvilinear coordinates
in the Euclidean space $\MR^3$ will be denoted by 
Greek letters $\,x^\mu$, $\mu=1,2,3$. Then the
``induced'' metric for the tube dislocation in 
cylindrical coordinates is\footnote{Note1}
\begin{equation}                                                  
\label{eincym}
g_{\mu\nu}(x)\,=\,\frac{\pl y^\rho}{\pl x^\mu}\,
\frac{\pl y^\s}{\pl x^\nu}\,\,\overset{\circ}g_{\rho\s}\,,
\end{equation}
where $\overset{\circ}g_{\rho\s}$ is the Euclidean 
metric written in cylindrical coordinates. We denote 
cylindrical coordinates of a point before the dislocation
is made by $\lbrace y,\vf,z\rbrace$, where $y$ without 
index stands for the radial coordinate and we take into 
account that the coordinates $\vf$ and $z$ do not 
change. Then the diffeomorphism is described by a 
single function relating old and new radial coordinates 
of a point $y=r-u(r)$, where
\beq
u(r) = \begin{cases} ar\,,\quad &r<r_*
\\ -\displaystyle\frac br\,,\quad & r>r_*\,. 
\end{cases}
\label{edifun}
\eeq
It is easy to see that this function has a discontinuity
$ar_*+b/r_*=l$ at the point of the cut. Therefore a special 
care must be taken in calculating the components of induced 
metric. It proves useful to introduce the function
\beq
v = \begin{cases} a\,,\quad & r\le r_*\,, 
\\ \displaystyle\frac b{r^2}\,,\quad & r\ge r_*\,,
\end{cases}
\eeq
which is continuous on the cutting surface. This function 
differs from the derivative $u^\prime(r)$ of the displacement 
vector field $u(r)$ defined in (\ref{edifun}) by the 
$\dl$-function
\beq
u' = u^\prime(r) = v(r) - l\dl(r-r_*)\,.
\label{displace}
\eeq
Now we define the induced metric for the tube dislocation as
\beq
ds^2 = (1-v)^2dr^2 + (r-u)^2d\vf^2 + dz^2\,.
\label{eindum}
\eeq
It is easy to check that this metric outside the cut agrees 
with the expression (\ref{eincym}). The difference between the 
two expressions is that (\ref{eindum}) is defined on the 
cutting surface while (\ref{eincym}) is not.
The volume element corresponding to (\ref{eindum}) is
\begin{equation*}
\vol=(1-v)(r-u)\,,\quad \text{where} \quad 
g=\det g_{\mu\nu}\,.
\end{equation*}
Let us note that the metric (\ref{eindum}) differs from the 
one which results from the formal substitution of $\,y=r-u(r)\,$ 
into the Euclidean metric $\,ds^2=dy^2+y^2d\vf^2+dz^2\,$ by the 
square of the $\,\dl$-function in the $\,g_{rr}$ component. This 
procedure is required in the geometric theory of defects, 
because otherwise the Burgers vector can not be expressed as 
the surface integral \cite{Katana05}. Finally, the metric 
component $\,g_{rr}(r)=(1-v)^2\,$ of tube dislocation is a 
continuous function, and the angular component 
$\,g_{\vf\vf}=(r-u)^2\,$ has the jump at the surface of the cut.

The displacement vector field (\ref{displace}) and induced 
metric (\ref{eindum}) represent the solution of the field 
equations of the linear elasticity theory (\ref{eqsepu}) 
with the boundary conditions (\ref{ebotud}) describing tube 
dislocation. The solution does not depend on the elastic 
properties of the media due to the universality of the 
linear approximation.

\section{Tube dislocation in the geometric theory of defects}

Our next step is to go beyond the linear approximation and 
consider the defects in elastic media with a tube dislocation. 
In the geometric theory of defects 
\cite{KatVol92,KatVol99,Katana03,Katana04} 
(see \cite{Katana05} for the for review) the defects in 
elastic media with a spin structure are the framework of 
differential geometry. The main assumption is that the 
elastic media is a three-dimensional manifold with a 
Riemann-Cartan geometry. This approach enables one to 
treat the non-linear regime and also enables one to deal 
with a continuous distribution of defects. 
The tube dislocation in the geometric theory of defects was 
introduced and considered in details in Ref. \cite{deBKat09}, 
so here we present only a short account of the formalism.  

The geometry is determined by the Einstein´s equations with 
the source term given by the energy-momentum tensor. We 
remark that the formalism under discussion is diffeomorphism 
invariant and thus, in contrast to the linear elasticity 
theory, the displacement vector field does not show up 
explicitly in the field equations. At the same time we 
require that the linear approximation gives standard result. 
Therefore a natural choice for the energy-momentum tensor is 
\begin{equation}                                              
\label{enmode}
  -\frac12T_{\mu\nu} = 
  \vol\left(\widetilde R_{\mu\nu}-
  \frac12g_{\mu\nu}\widetilde R\right)\,,
\end{equation}
where the curvature terms are constructed by using the 
metric obtained within the elasticity theory, e.g., 
(\ref{eindum}) for the tube dislocation case. The 
straighforward calculations yeild the following
relevant component of the above equation \cite{deBKat09}:
\begin{equation}                                                  
\label{etusof}
T_{zz}\,=\,\frac{4lr_*}{2r_*-l}\,\,\delta^\prime(r-r_*)\,.
\end{equation}
It is worth mentioning that the above calculation is 
non-trivial because the curvature terms contains 
ambiguous pieces, like $\delta^2$, but surprisingly 
they cancel in the final expression.

It turns out that the solution in the geometric theory of 
defects in some appropriate gauge conditions coincides 
exactly with the metric (\ref{eindum}). Thus, the result 
obtained within the linear elasticity theory is quite 
general and we can adopt the corresponding geometry as 
a background for the quantum consideration.

\section{Schr$\ddot{\rm o}$dinger equation in the 
presence of tube defect}

Consider the quantum description of a spinless particle 
in the space with tube dislocation. The quantum properties 
of our interest are encoded into the time independent 
wave function, $\psi(r,\phi,z)$  which is the solution 
of the stationary covariant Schr$\ddot{\rm o}$dinger equation,
\beq
\Delta \psi = -\chi^2\,\psi\,.
\eeq 
Here we use the notation $\chi = \sqrt{2mE}/\hbar$, where
$m$ and $E$ are mass and energy of the particle. $\Delta$ 
is the covariant Laplacian operator, given by
\beq
\Delta \psi \,=\, 
\frac{1}{\sqrt{g}}\pa_\mu\,(\sqrt{g}g^{\mu\nu}\pa_\nu \psi)\,.
\label{Delta}
\eeq
Notice that the wave function behaves as a scalar field 
when subject to covariant derivative, because the particle 
is spinless. For this reason (\ref{Delta}) boils down to 
eq. (\ref{cyl}) in cylindrical coordinates and, furthermore,
the torsion does not couple with the wave 
function
\footnote{Note2}. 
The straightforward calculations lead to the following 
form of Schr$\ddot{\rm o}$dinger equation in the spacetime 
with the elastic deformations described in section 2,
\beq
\frac{1}{(1-v)^2}\frac{\pa^2\psi}{\partial r^2} 
+ \frac{1}{(1-v)^2}\left(\frac{1-u^\prime}{r-u} 
+ \frac{v^{\prime}}{1-v}\right)\frac{\pa\psi}{\pa r} 
+ \frac{1}{(r-u)^2}\frac{\pa^2\psi}{\pa \phi^2} 
+ \frac{\pa^2\psi}{\pa z^2}
= -\chi^2 \psi\,.
\label{Schrodinger}
\eeq
In order to obtain the energy spectrum and density 
distribution, we need to to solve the above equation 
separately in the two distinct regions, namely for 
\ $r < r_\ast$ \ (this implies also $y<r_1$) \ and 
for \ $r > r_\ast$ (in this case $y>r_2$). We shall 
denote the corresponding solutions for the wave 
functions as \ 
$\psi^{{\rm int}}(r)$ \ and \ $\psi^{{\rm ext}}(r)$. 
One has to take into account also the normalization 
and boundary conditions, along with the matching 
condition at \ $r=r_\ast$, namely,
\beq
\psi^{{\rm int}}(r_\ast) \,=\, \psi^{{\rm ext}}(r_\ast)\,,
\eeq
\beq
\Big(\frac{\pa\psi^{{\rm int}}}{\pa r}\Big)_{r=r_\ast} 
\,=\, 
\left(\frac{\pa\psi^{{\rm ext}}}{\pa r}\right)_{r=r_\ast}\,.
\label{match}
\eeq
It would be very difficult to solve the equation 
(\ref{Schrodinger}) in a direct way, especially outside 
the tube defect. However the solution becomes rather 
straightforward if we perform the inverse coordinate 
transformation $\,r\to y = r - u(r)$. In this case
the Schr$\ddot{\rm o}$dinger equation becomes very simple and in 
fact it is just the one for the free particle,
\beq
\frac{\pa^2\psi}{\pa y^2} 
+ \frac{1}{y}\frac{\pa\psi}{\pa y} 
+ \frac{1}{y^2}\frac{\pa^2\psi}{\pa \phi^2} 
+ \frac{\pa^2\psi}{\pa z^2} 
= - \chi^2 \psi\,. 
\label{Schrodinger2}
\eeq
After the equation (\ref{Schrodinger2}) is solved one 
has to perform the coordinate mapping, that means the 
direct elasticity transformation in order to get the 
solution of the equation (\ref{Schrodinger}) of our 
interest. It is important to keep in mind that the 
triviality of the equation (\ref{Schrodinger2}) does 
not mean the triviality of the solutions of the 
equation of our interest (\ref{Schrodinger}). The 
whole point is that the solutions are coordinate 
dependent (while the equation is covariant) and the 
physical results correspond to the coordinates $x^i$
defined in eq. (\ref{eeldef}).

The solution of the free Schr$\ddot{\rm o}$dinger equation 
(\ref{Schrodinger2}) can be easily obtained via the 
usual method of separating variables if we consider 
a finite cylinder between $z=-L$ and $z=L$. We set 
$\psi (y,\phi ,z) = \rho (y) \Phi (\phi) Z(z)$ and 
arrive, after replacing this form into the equation 
and some standard algebra, at the solutions for 
$\phi$ and $z$ dependences,
\beq
\Phi(\phi) = e^{im\phi} 
\quad {\rm and} \quad 
Z(z) = \cos (kz)\,,
\eeq
where $m = 0,1,2,...$ and $k = n\pi/L$ ($n=1,3,5,...$). 
These solutions correspond to the boundary conditions 
$Z(-L)=Z(L)=0$. The radial function satisfies the equation
\beq
\frac{\pa^2\rho}{\pa y^2} + \frac{1}{y}\frac{\pa\rho}{\pa y} 
\,=\, 
-\,\Big(\bar{\chi}^2 - \frac{m^2}{y^2}\Big)\rho\,. 
\label{radial eq}
\eeq
where $\bar{\chi}^2 = \chi^2 - k^2$. 
The general solution for the last equation has the form
\beq
\rho(y)\, = \,C_1\, J_m(\bar{\chi} y) 
\,+\, C_2 \, Y_m(\bar{\chi} y)\,,
\label{general sol}
\eeq
where $J_m$ and $Y_m$ are Bessel function of the first 
kind and of the second kind (or Neumann function) 
respectively, and $\,C_1,\,C_2\,$ are arbitrary constants.
The above solution can be mapped from the ``old'' 
coordinate $y$ into physical coordinate $r$,
as it was explained above, providing us an important 
information on the spectrum of the system and also on 
the charge density distribution. Let us discuss this in 
some details. 

\subsection{Energy levels}

Let us consider the solution of Schr$\ddot{\rm o}$dinger 
equation for a particle confined in a thin region around 
the geometrical defect, say, in old coordinates, 
\ $r_0\leq y\leq r_1$, \ $r_2\leq y\leq r_3$. The two 
separate solutions are
\beq
\rho^{int}(y) 
& = & c_1 J_m(\bar{\chi} y) + c_2 Y_m(\bar{\chi} y)\,,
\;\;\; y < r_1\,, \\
\rho^{ext}(y) 
& = & c_3 J_m(\bar{\chi} y) + c_4 Y_m(\bar{\chi} y)
\,,\;\;\; y > r_2\,.
\eeq
After the necessary mapping, $r = y + u$, we arrive at 
the exact solutions of the radial part of the equation 
(\ref{Schrodinger}). 
One should mention that a real finite-size tube does not 
have the symmetry along the $z$-axis, however one can 
consider a very long and thin tube and thus assume the 
translational $z$-symmetry as a good approximation. In 
our case this means an approximation of a large $L$, such 
that $k\simeq 0$. 

Consider the mapping, which was already discussed above, 
in the detailed form. In what follows, we shall drop the 
bars and write $\chi$ instead of $\bar{\chi}$. In order to 
write down the solution in terms of physical coordinate, 
we have to impose the boundary conditions in the form
\beq
c_1 J_m(\chi r_0) + c_2 Y_m(\chi r_0) = 
c_3 J_m(\chi r_3) + c_4 Y_m(\chi r_3) = 0\,.
\label{boundary}
\eeq 
These conditions give us, immediately, 
\ $c_2 = -c_1 J_m(\chi r_0)/Y_m(\chi r_0)$ \ and 
\ $c_4 = -c_3 J_m(\chi r_3)/$ \\ $Y_m(\chi r_3)$. The last 
relations must be inserted on the matching condition 
at $r=r_\ast$. However, it turns out that the old 
matching conditions (\ref{match}) are physically
unacceptable, since they imply that the charge density,
proportional to $\sqrt{g} |\psi |^2$, has a discontinuity 
at the surface $r_\ast$. Thus, the following conditions 
would be more appropriate:
\beq
r_1 \left|\rho(r_1)\right|^2 
\,=\, r_2 \left|\rho(r_2)\right|^2 \,,
\qquad
\left.\frac{d}{dy}\,\left(y |\rho(y)|^2\right)\right|_{y=r_1} 
\,=\, 
\left.\frac{d}{dy}\,\left(y |\rho(y)|^2\right)\right|_{y=r_2}\,.
\label{matching2}
\eeq
Here we have used the expression for the volume element 
in cylindric coordinates, $\sqrt{g(r)}=r$. 

Unfortunately, the above conditions provide only two 
relations between arbitrary (complex) constants, in contrast
to the equations (\ref{match}), which provide four relations.
Thus one should use another conditions (complex equations).
The most natural choice is continuity 
of the covariant density $\,\sqrt{g}|\psi|^2$. 
Another condition with direct
physical interpretation is the current conservation, that 
is the continuity of the current \ $\sqrt{g}J^\mu$, 
that means \ $\nabla_\mu J^\mu=0$, 
where
\beq
J^\mu \,=\, -\,\frac{i\hbar}{2\pi}\,g^{\mu\nu}
\big(\psi^\ast\pa_\nu\psi - \psi\pa_\nu\psi^\ast\big)\,.
\nonumber
\eeq 
It turns out that the relations (\ref{matching2}) or,
in a more general form, 
\beq
g^{1/4}(r_1)\rho (r_1) 
& = & 
g^{1/4}(r_2)\rho (r_2) \,,
\nonumber 
\\
\nonumber
\\
g^{1/4}\,g^{\mu\nu}\,\pa_\nu\rho(y)\Big|_{y=r_1} 
& = &
g^{1/4}\,g^{\mu\nu}\,\pa_\nu\rho(y)\Big|_{y=r_2}\,,
\label{matching4}
\eeq
imply the above physical assumptions and provide four
necessary relations. On the top of that we have the 
normalization condition 
\beq
\int_V\,\big|\psi\big|^2 dV \,=\, 1\,.  
\label{normalization}
\eeq

The equations (\ref{matching4}) 
determine the discrete energy spectrum. Of course the 
results are strongly dependent on the values of parameters
$r_0$, $r_1$, $r_2$ and $r_3$. In order to compare the 
results with the available data, let us 
us specify these parameters as follows: 
\beq
r_0 = 10\,nm\,, \quad
r_1 = 54\,nm\,, \quad
r_2 = 56\,nm \quad \mbox{and} \quad
r_3 = 100\,nm.
\label{nm}
\eeq 
Typically, the outer diameter of CNT ranges between 2 and 
20 nm and inner diameter ranges between 1 and 3 nm,
interlayer distance is 3.4 nm as it is established by 
the high-resolution TEM techniques.
The values (\ref{nm}) correspond to the typical spacing 
between the layers for nanotubes, confirmed, in particular, 
by the experiments using diffraction and HRTEM image 
techniques \cite{nm1,nm2}. We consider here the possibility
of describing MWNT, with several layers of carbon or other 
materials. Indeed, since we are looking for a qualitative 
correspondence with the results obtained by other methods, 
there is no much sense in performing calculations for large 
amount of possible nanotube diameters. However, since our 
method is really simple, such calculation can be 
easily performed, e.g., for any particular MWNT where
the fast theoretical evaluation of the effect is 
needed. 
The system,  as showed in Figure \ref{fig3}, can be 
treated geometrically, by identifying nanotubes with 
geometrical properties. In our consideration, a 
different kind of nanotube at $r = 55$ $nm$ means a 
geometrical defect at the corresponding region.

Let us note that, taking the radius magnitude close to 
the ones which are typically found in nanosystems, 
including MWNT's, we also assume that the (e.g. carbon) 
nanotubes layers are of the same chirality. Indeed, 
one can suppose that a nanotube located at the middle, 
at $r = 55$ $nm$, has different chirality, as well as, 
pehaps, few nanotubes in its vicinity \cite{chir}.

\begin{quotation}
\begin{figure}
\includegraphics[scale=0.5]{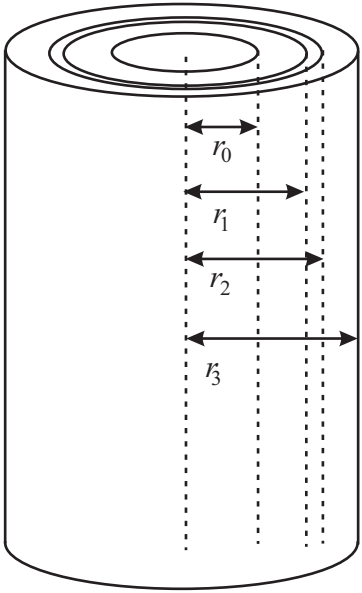}
\caption{An illustration of the general geometrical 
form of a MWNT with the inner diameter of 20 $nm$ 
and outer diameter of 200 $nm$. \label{fig3}}
\end{figure}
\end{quotation}

Let us find the energy spectrum for the above choice, which
depends on the quantum number $m$. We shall start from the 
fundamental state, $m=0$.
One can write equation (\ref{matching4}) in terms of just
one arbitrary constant, say, $c_1$, which can be fixed by 
using the condition (\ref{normalization})). The corresponding
equation is transcendental and rather cumbersome, so it is 
not convenient to write it here. The solution was performed 
with the help of the software {\it Mathematica} \cite{M6}. 
We have found the values for $\chi = \sqrt{2mE}/\hbar$ which 
are solutions of eq. (\ref{matching4}). The first three 
eigenvalues are placed in the left column of Table 1. 
For the purpose of comparison, we put the 
corresponding eigenvalues for the simplest case of the 
condicting cylinder without geometrical defect, in the 
right column. Similar results are presented in Table 2 
for the $m=1$ case.

\begin{center}
\begin{tabular}{c||c|c}
		$\chi$ ($nm^{-1}$) & with defect ($l = 2\,nm$) 
		& without defect ($l = 0\,nm$) \\ \hline\hline
		$\chi_1$ & 0.03383 & 0.03314 \\ \hline
		$\chi_2$ & 0.07016 & 0.06858 \\ \hline
		$\chi_3$ & 0.10612 & 0.10377 \\ \hline
\end{tabular}
\end{center}
\begin{center}
Table 1: Energy spectrum for $m=0$ with tube defect (left)
and without defect (right). The quantities $\chi_1$, $\chi_2$ 
and $\chi_3$ correspond to the first three energy eigenvalues.
\end{center}

From these results, one can see that the tube defect does
shift, at the first place, the energy spectrum as a whole 
while it almost does not change the energy gap. Still there 
is a very small effect and the tendency is to reduce the 
gap slightly in all cases. Similar situation occurs
for $m=1$ (see Table 2). This shift depends on
the size $l=r_2 - r_1$ of the cutting region. In Table 3,
we put the first three energy eigenvalues for the cases
$l=0\,nm$, $l=-2\,nm$, $l=2\,nm$ and $l=4\,nm$, for $m=0$. 
Of course, $l=0$ corresponds to the case without defect. 
The case $l=-2\,nm$ corresponds to a negative tube dislocation 
(see Figure 1b).

\begin{center}
\begin{tabular}{c||c|c}
		$\chi$ ($nm^{-1}$) & with defect ($l = 2\,nm$) & 
		without defect ($l = 0\,nm$) \\ \hline\hline
		$\chi_1$ & 0.04009 & 0.03941 \\ \hline
		$\chi_2$ & 0.07490 & 0.07331 \\ \hline
		$\chi_3$ & 0.10982 & 0.10748 \\ \hline
\end{tabular}
\end{center}
\begin{center}
Table 2: Energy spectrum for $m=1$ with tube defect (left)
and without defect (right). 
\end{center}

It is remarkable that the energy levels do shift under 
the mapping from the ``old'' coordinates to the physical 
ones while one could naively expect that these levels 
remain the same. The reason is that (as we have already 
mentioned) the mapping does not change the {\it general}
solutions of the Schr$\ddot{\rm o}$dinger equations, 
while our interest is concentrated on the particular 
solutions with the given boundary conditions. These 
conditions do change under the mapping and this makes 
the solutions of the equation (\ref{Schrodinger}) 
really nontrivial. 

\begin{center}
\begin{tabular}{c||c|c|c|c}
		$\chi$ ($nm^{-1}$) & $l = -2\,nm$ & $l = 0\,nm$ & 
		$l = 2\,nm$ & $l = 4\,nm$ \\ \hline\hline
		$\chi_1$ & 0.03248 & 0.03314 & 0.03383 & 0.03456 \\ \hline
		$\chi_2$ & 0.06706 & 0.06858 & 0.07016 & 0.07183 \\ \hline
		$\chi_3$ & 0.10153 & 0.10377 & 0.10612 & 0.10858 \\ \hline
\end{tabular}
\end{center}
\begin{center}
Table 3: The three first energy levels for $m=0$ for the 
cases $l = -2\,nm$, $l = 0\,nm$, $l = 2\,nm$ and $l = 4\,nm$. 
\end{center}

How is the dependence of the energy levels
on the size $l$? As suggested by the data of Tables 1, 2 and 3, 
positive $l$ induces a slight increase of energy 
eigenvalues, while a negative $l$ lowers them. The
exact function $\chi_i (l)$ ($i=1,2,3$) is unknown
and all we can do is to plot the data from Table 3
as shown in Figure \ref{fig0}. This plotting clearly 
suggests a linear dependence between the energy levels 
$\chi$ and the size $l$.

\begin{quotation}
\begin{figure}
\includegraphics[scale=0.7]{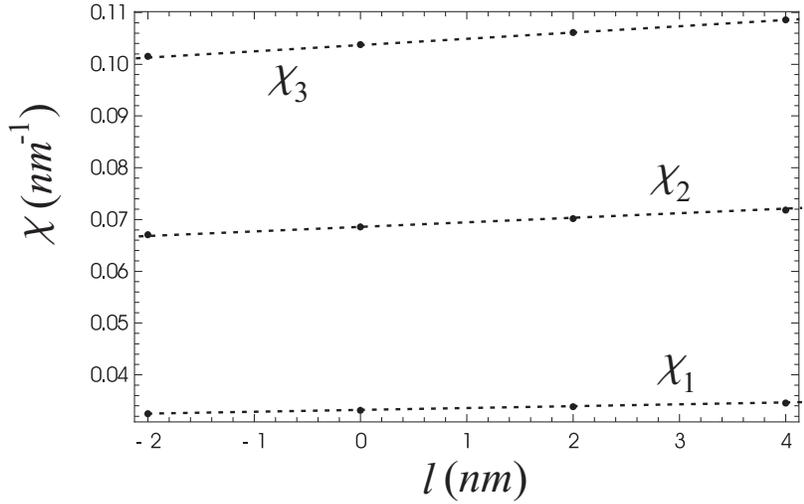}
\caption{Representation of the data of Table 3 in the plane
$\chi\times l$. The first three eigenvalues, 
$\chi_1$, $\chi_2$ and $\chi_3$
are plotted for the cases $l = -2\,nm$, $l = 0\,nm$, $l = 2\,nm$ 
and $l = 4\,nm$. The dashed lines are supposed to 
describe the assumed linear dependence. \label{fig0}}
\end{figure}
\end{quotation}

One can notice, by straighforward computation, that 
the energy gaps, $\Delta \chi_1 :=\chi_2 - \chi_1$
and $\Delta \chi_2 :=\chi_3 - \chi_2$, are slightly
increased for $l>0\,nm$ and decreased for $l<0\,nm$. For example,
for $l=4$ $nm$, $\Delta \chi_2$ is 4.4\% greater than
the corresponding value of $\Delta \chi_2$ for $l = 0\,nm$.
The values of $\Delta \chi_{1,2}$ are shown in Table 4,
and are plotted in Figure \ref{fig8}. We can observe again 
a linear relation between $\Delta \chi$ and $l$. Let us notice
that, by assuming linear dependence, the straight line
describing $\Delta \chi_1 (l)$ has a lower angular
coefficient than the one corresponding to $\Delta \chi_2 (l)$.

\begin{center}
\begin{tabular}{c||c|c|c|c}
		$\Delta \chi$ ($nm^{-1}$) & $l=-2\,nm$ & $l=0\,nm$ & 
		$l=2\,nm$ & $l=4\,nm$ \\ \hline\hline
		$\Delta \chi_1$ & 0.03458 & 0.03544 & 0.03633 & 0.03727 
\\ \hline
		$\Delta \chi_2$ & 0.03447 & 0.03520 & 0.03596 & 0.03675 
\\ \hline
\end{tabular}
\end{center}
\begin{center}
Table 4: The first two energy gaps, 
$\Delta \chi_1 = \chi_2 - \chi_1$ and 
$\Delta \chi_2 = \chi_3 - \chi_2$, 
for $m=0$ for different values of $l$. 
\end{center}

\begin{quotation}
\begin{figure}
\includegraphics[scale=0.7]{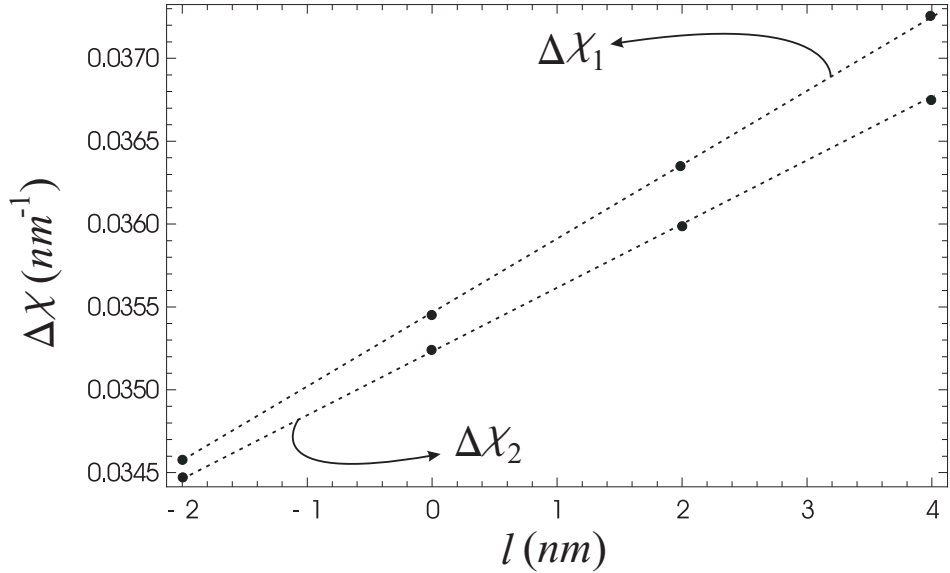}
\caption{Representation of the data of Table 4 in the plane 
$\Delta \chi\times l$.
The two first energy gaps, $\Delta \chi_1 = \chi_2 - \chi_1$ 
and $\Delta \chi_2 = \chi_3 - \chi_2$, are drawn for
different values of $l$. The dashed lines are supposed to 
describe the assumed linear dependence. \label{fig8}}
\end{figure}
\end{quotation}

\subsection{Density Curves} 
 
In the ideal gas approximation, the volume integral 
of the density distribution, \\
$\int\,\sqrt{g}|\psi(y,\phi,z)|^2\,d\phi dz dr$,
is proportional to $\int\,y|\rho(y)|^2\,dr$, with the
constant of proportionality having dimensions of area. From the 
condition (\ref{normalization}), $|c_1|^2$ has dimensions of 
[length]$^{-3}$ and one can find (using 
{\it Mathematica} software) the following values for $|c_1|^{-2}$:
\ $2265.29$ $nm^{3}$, \ 
$8985.42$ $nm^{3}$ and $8151.17$ $nm^{3}$ for $\chi_1$, 
$\chi_2$ and $\chi_3$, respectively,
for $m=0$, with tube defect. The corresponding values without
defect, for $\chi_1$, $\chi_2$ and $\chi_3$ are 
$2315.70$ $nm^{3}$, $8017.33$ $nm^{3}$ and $11471.10$ 
$nm^{3}$, respectively.

With these values, one can plot the normalized density against
the coordinate $y$. In Figure \ref{fig1}, for the lowest 
energy eigenvalue, $\chi_1$, and $m = 0$, the normalized density
is plotted as well as the normalized density without defect for
comparison. Notice that the curve in the presence of defect has
a cut because the point $r_1$ is identified with $r_2$, so the 
interval $r_1 < y < r_2$ is formally absent in the space with
tube defect. The defect raises density in the central region.
It is interesting to observe that this is not the unique effect
caused by the defect. With the help of the {\it Mathematica}
software, one can calculate the value of $y$ for which the
density reaches its maximum value. This value is given by
$y_{{\rm max}}\approx 53.80\, nm$ for the density without
defect and $y_{{\rm max}}\approx 52.95\, nm$ for the
density with $m=0$ and $l = 2\,nm$. Thus, the defect not only 
raises the density distribution, but it also drags its peak
to the left.

Let us remark that the curves describing the particle density
have not symmetry around the straight line $y = 55\,nm$. The 
peaks of the curves in Figure \ref{fig1} are localized in the 
left half of the region $10\,nm \leq y \leq 100\,nm$, and not 
in the middle of it. Actually, this property is essentially 
related to the properties of the Bessel functions. Indeed, 
let us remmember that the system is not unidimensional, such that
it does not possess symmetry around the middle point. 

\begin{quotation}
\begin{figure}
\includegraphics[scale=0.6]{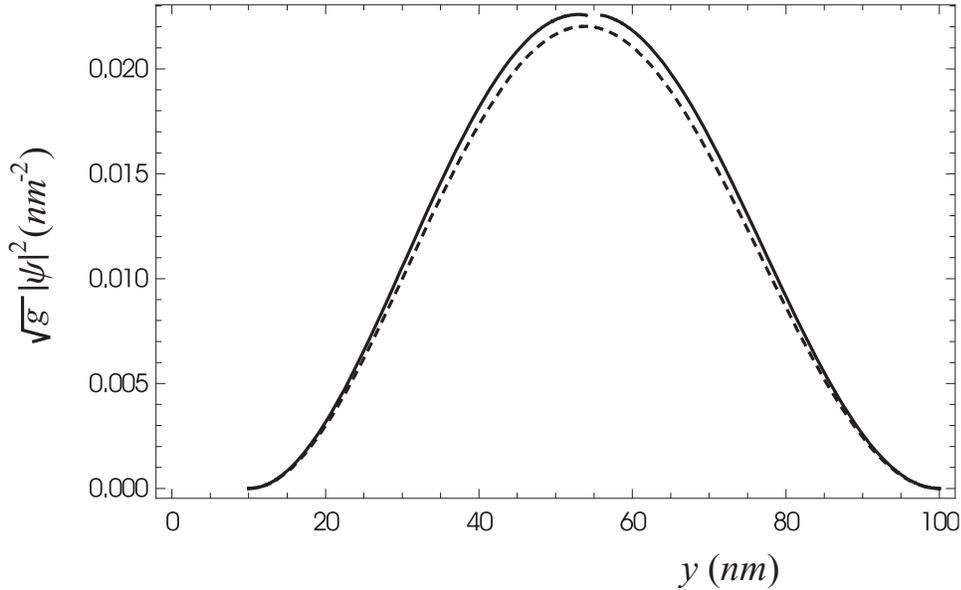}
\caption{Normalized density for $m=0$ and $\chi = \chi_1$. 
The continuous line has a cut and describes density
in the presence of tube defect. The dashed line describes density
without defect. \label{fig1}}
\end{figure}
\end{quotation}

For $\chi = \chi_2$ and still $m = 0$, the corresponding density
curves are plotted in Figure \ref{fig2}. In the presence of the 
tube defect, the density is lower in central region but greater 
in the marginal regions, $y \approx 30\,nm$ and $y \approx 80\,nm$. 
The shifts in the density distribution
are small and of the same order in both cases, $\chi_1$ and 
$\chi_2$, but in the last case the difference is richer. In the 
central region, the defect raises the density for $\chi = \chi_1$ 
and lowers the density for $\chi = \chi_2$. Of course, for a larger
$l$, these effects may become much more sensitive.

\begin{quotation}
\begin{figure}
\includegraphics[scale=0.9]{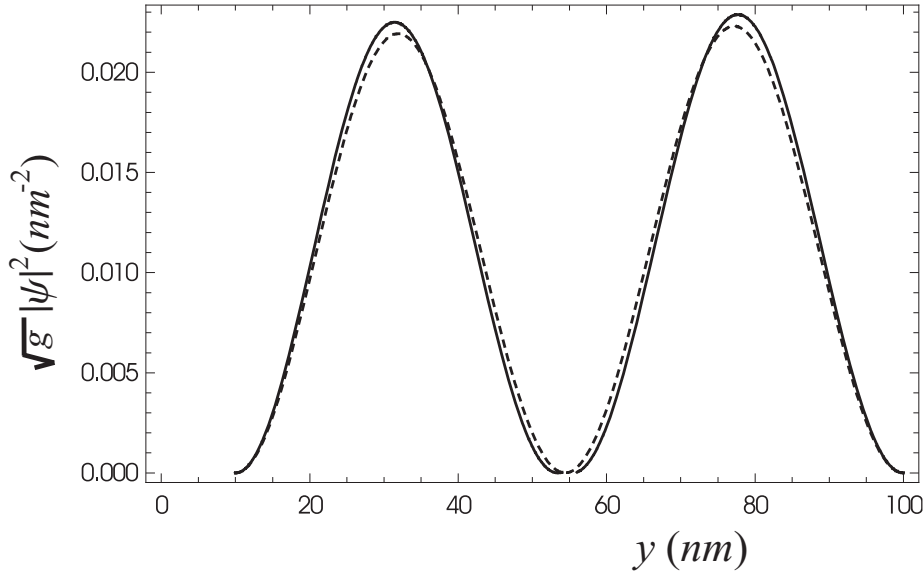}
\caption{Normalized density for $m=0$ and $\chi = \chi_2$. 
The continuous line has a cut and describes density
in the presence of tube defect. The dashed line describes density
without defect. \label{fig2}}
\end{figure}
\end{quotation}


\section{Conclusions}

We considered the solution of the Schr$\ddot{\rm o}$dinger 
equation in the restricted region of space with cylindrical 
geometric defect. The region is confined between the two 
cylindrical shells and the diameter of the defect is 
intermediate between the ones of these shells. 
Geometrically this configuration resembles the double-wall 
nanotube (DWNT) and, therefore, it can be regarded as a simplified 
model of such nanotube. Due to the fact that the electron 
can freely move within the space between the two border 
shells, one can associate the system with the M/M-type 
DWNT. The solution of the Schr$\ddot{\rm o}$dinger equation 
shows the influence of the geometric defect on the energy 
gap and charge distribution. In particular, one meets a 
modified energy spectrum and the distribution of charge 
density, in the ideal gas approximation. These effects 
are qualitatively similar to the ones previously reported for 
some MWNTs in \cite{gap}, where the calculations were based 
on the tight-binding method. In this respect we can conclude
that the approach based on geometric defects, which does 
not take into account anything but the general form of the 
DWNT, is able to provide some (very restricted, indeed) 
information about the electron behaviour. 

It is obvious that the method based on geometric defects 
can not compete with the standard approaches based on 
molecular dynamics. The reason is that the geometric 
defects method can not take into account full details 
of the structure of the compound and is, in some sense, 
too general. At the same time this method may become much 
more interesting if one develops it further and, in particular, 
learns how to deal with more sophisticated versions of 
geometric defects. In particular, it looks possible to 
take into account the chirality of the nanotube and, 
also, include the external magnetic field. We expect to 
consider these issues elsewhere.


\section*{Acknowledgments}
G.B.P., E.K. and I.Sh. are indebted to CNPq, FAPEMIG and 
FAPES for partial support. M.K. is thankful to FAPEMIG for 
support of his visit to Brazil and to the Physics Department 
of the Universidade Federal de Juiz de Fora for warm 
hospitality. Also, he thanks the Russian 
Foundation of Basic Research (Grant No.\ 08-01-00727), and
the Program for Supporting Leading Scientific Schools
(Grant No.\ NSh-3224.2008.1) for financial support.
The work of I.Sh. has been also supported by ICTP.


\end{document}